# Uncovering patterns of inter-urban trip and spatial interaction from social media check-in data

Yu Liu, Zhengwei Sui, Chaogui Kang, Yong Gao
Institute of Remote Sensing and Geographical Information Systems, Peking University, Beijing 10081, China

Abstract: The article revisits spatial interaction and distance decay from the perspective of human mobility patterns and spatially-embedded networks based on an empirical data set. We extract nationwide inter-urban movements in China from a check-in data set that covers half million individuals and 370 cities to analyze the underlying patterns of trips and spatial interactions. By fitting the gravity model, we find that the observed spatial interactions are governed by a power law distance decay effect. The obtained gravity model also well reproduces the exponential trip displacement distribution. However, due to the ecological fallacy issue, the movement of an individual may not obey the same distance decay effect. We also construct a spatial network where the edge weights denote the interaction strengths. The communities detected from the network are spatially connected and roughly consistent with province boundaries. We attribute this pattern to different distance decay parameters between intra-province and inter-province trips.

## 1. Introduction

A number of social media websites that support geo-tagged information submission and sharing have been recently introduced and achieved great commercial success. Various functions have been provided by these websites, such as social network (Facebook), micro-blog (Twitter), photo sharing (Flickr), and location based check-in (Gowalla and Foursquare). Each website has millions of registered members and their submissions form an important type of big data. Since much information is associated with particular locations, Goodchild coined the term volunteered geographical information (VGI) for it [1]. In this paper, we use the term "check-in record" for a piece of geo-tagged content posted by a user. A check-in record generally includes a short textual message, a photo, and the time and location indicating when and where the message was posted. With a check-in data set, we can extract the footprints of large volumes of individuals. Although the trajectory of one particular person is rather stochastic, we can find underlying patterns when the number of trajectories gets greater. An interesting example is a map depicting the last 500 million check-in points on Foursquare that clearly demonstrate the human activity distribution across the world[1]. Much research has been conducted using check-in data, sometimes with additional data such as social ties between users, collected from various sources. Several strands of status quo work can be identified. At the individual level, human mobility patterns [2,3] and geographical impacts on social networks [4,5] are investigated. At the aggregate level, these data enables us to study spatial activity distributions and spatial interactions between regions [6].

Recently, human mobility patterns have drawn much in the areas of physics [7], geography [8,9], and computer science [10], with the availability of multi-sourced trajectory data. However, these studies either do not distinguish motion patterns at different spatial scales or focus on intra-urban trip patterns. It is natural that inter-urban trips have different mechanisms from those of intra-urban trips. For example, one in general has two frequently revisited anchor points (i.e. home and workplace) and commutes occupy a large proportion in intra-urban trips. On the contrary, we can only find one anchor point, corresponding to his (or her) home town, from an individual's trajectory at the inter-city scale. However, whether there exists different mechanisms account for different human mobility patterns at and across different scales remains a research question. Little comparison research on this point has been done due to the lack of individuals' inter-urban trajectories. Clearly, a check-in data set makes an investigation of inter-urban mobility possible for its large spatio-temporal coverage.

In this research, we use a social media check-in data set submitted by about half millions users to study the inter-urban trip patterns. At the collective level, these trips represent spatial interaction strengths between cities. Our research serves three purposes. First, we intend to reveal the underlying distance effect in the trips extracted from check-in records. Second, we try to link patterns at the collective level of spatial interactions versus the individual level of human movements, and to make a comparison with intra-urban patterns revealed from mobile phone or taxi data sets. Last, we investigate the implications of distance decay effect in regionalizing the study area based on spatial interactions between cities.

## 2. Background

This section summarizes research in three areas: spatial interaction, human mobility pattern, and spatially-embedded network. The first is a fundamental topic in geographical applications, and the last two have recently drawn much attention in both geographical and physical studies, with the availability of spatio-temporally-tagged big data. This research reveals the underlying connections among them using empirical data set.

### 2.1 Distance decay effect in spatial interactions

Spatial interactions between geographical entities such as cities and regions help us to understand spatial structure of a region and plan an efficient spatial configuration. In practice, interaction strength can be measured by volumes of passengers [11], migration flows [12], trade flows, currency flows, telecommunications [13-15], or even toponym co-occurrences [16]. Due to the complexity of spatial interaction involving pairs of multiple spatial nodes, much research has also been conducted on effectively visualizing spatial interactions and delineating meaningful sub-regions [17,18].

---

[1] https://foursquare.com/infographics/500million.

Most spatial interaction systems are governed by the distance decay effect [19,20], which is in general expressed in the gravity model [21]. Derived from Newton's law of gravity, the gravity model in geographical applications is formulated as $I_{ij} = \frac{kP_iP_j}{f(d_{ij})}$, where $I_{ij}$ and $d_{ij}$ denote the interaction from *i* to *j* and distance between two places, and $P_i$ and $P_j$ are repulsion of place *i* and attraction of place *j*, respectively. If we do not distinguish the two directions, $I_{ij}$ denotes the sum of flows from *i* to *j* and *j* to *i*. Meanwhile, $P_i$ and $P_j$ are often approximated by the sizes of the places. The gravity model has been widely used for estimating traffic and migration flows. In the model, distance decay function can vary with applications of interest (e.g., traffic flow versus migration) and technological renovation [22], and one fixed decay function might not fit all problems. Population size might not be able to accurately describe the ability of repulsion or attractiveness of places. A number of studies have then used the observed interaction strengths and distances between geographical entities to fit the gravity model, resulting in the theoretical size (or nodal attraction) of each entity and the distance friction function *f(d)*. Wang [23] summarized several forms of *f(d)*, among which the power law function $d^{-\beta}$ is widely used. The distance decay parameter β reveals the distance impacts on interaction behavior due to the scale free property of $d^{-\beta}$. We can compare different interaction behaviors using their β values. A greater β implies faster decay effect and the interactions are more influenced by distance.

A number of practical methods have been developed for fitting the gravity model, including linear programming [24] and simplified algebraic method [25,26]. Recently, the particle swarm optimization (PSO) method was introduced to fit the gravity model [11]. The merit of this method is twofold. First, it works well for interaction networks with low density, that is, the interactions of certain pairs of nodes are absent. Second, we can use different distance friction functions beyond the power law when optimizing the model to estimate the nodal attractions.

## 2.2 Human mobility patterns

Understanding human mobility patterns can help us in many fields including epidemic control and traffic management [27-29]. A number of data sources are introduced to study human mobility patterns. They include mobile phone call records [7,30], GPS (Global Positioning System) enabled taxi trajectories [8,9,31], smart card records in public transportation systems [32], and check-in data [2,3].

A number of measurements can be used to quantify human mobility patterns [33,34]. Among them, the distribution of displacements is extensively investigated. Existing studies reveal that the probability of a movement with distance *Δd*, denoted by *P(Δd)*, decreases with an increase of *Δd*, indicating the distance decay effect. Different studies suggest that *P(Δd)* can be fitted by different statistical distributions such as power law *P(Δd)~Δd$^{-\beta}$* [2,10], exponential law *P(Δd)~exp(-kΔd)* [30,31], or exponentially truncated power law *P(Δd)~exp(-kΔd)Δd$^{-\beta}$* [7,9]. The parameters in the above distributions are critical in applications such as epidemic or virus diffusion [28,35]. Particularly, when *P(Δd)* follows a power law distribution in which 1<β<3, and the direction distribution is uniform, the trajectory can be modeled by a Lévy flight.

Various models have been proposed to interpret the observed human mobility patterns. They takes into account different influencing aspects such as population characteristics [7], individuals' activities (e.g. returning to particular points, [36]), geographical environments [30,37-39], and distance effects [3,9,40]. These aspects are central to human geography such that the big-data-based human mobility research can shed light on understanding human environment interactions from a new perspective.

## 2.3 Spatially-embedded network

Given a set of geographical entities with known interaction strengths between them, we can construct a spatially-embedded network (or spatial network), in which each node is located in space so that the distance between each two nodes can be measured [41]. A spatial network may be tangible (e.g. street networks) or intangible (e.g. flight networks or networks constructed from social media). With the advances in complex network research, many geographical studies introduce complex network methods into geographical analyses [42-43].

In complex network analyses, detecting communities is an important task. Given a network, a community is a subset with relatively dense node-to-node connections. Many algorithms have been proposed for detecting communities, including Girvan-Newman method [45], multilevel method [46], fastgreedy method [47], infomap method [48], walktrap method [49], etc. In a community detection procedure, the modularity of a graph is widely used for measuring how good a division is. For a weighted graph, the modularity is computed as

$$Q = \frac{1}{2m} \sum_{i,j} \left( A_{ij} - \frac{k_i k_j}{2m} \Delta(c_i, c_j) \right) \quad (1)$$

where $m$ is the number of edges, $A_{ij}$ is the edge weight between nodes $i$ and $j$, $k_i$ and $k_j$ are the sum weights of edges linked to the two nodes. $c_i$ and $c_j$ denote the community of $i$ and $j$ and $\Delta(x,y)$ equals 1 when $x=y$ and 0 when $x \neq y$.

For a spatial network, a community corresponds to a region, which may be spatially connected or disconnected (i.e. with enclaves). Community detection methods are therefore extended to take into account specific spatial characteristics, such as adjacency constraint [17] and distance effect [50], for regionalization. However, there is some research directly uses conventional community detection methods for spatial networks, for example global flight network [51], telephone communication network [52,53], and network constructed from movements [54,55]. It is interesting that such networks yield spatially connected regions, and some regions coincide with administrative units rather well. For instance, De Montis et al. reported that the communities obtained from commuters' flows of Sardinia, Italy, in many cases match administrative configurations [54]. In this research, we try to interpret the spatial connectedness using the distance decay effect.

## 3. Data

## 3.1 Data description

This research uses a check-in data set collected from a major Chinese LBSNS (location-based social network service) provider, which can be viewed as a counterpart of Foursquare in the western world. The data set contains check-in records posted by approximately 521,000 registered users in one year, from September 2011 to September 2012[2]. Note that fake check-ins exist in the data set. A fake check-in record means that the distance between its real location, denoted by geographical coordinates, and its declared venue is greater than a threshold. After filtering out fake check-ins, we obtain about 23,500,000 records. The heat map of all check-in points clearly highlights the urbanized areas in China (Figure 1A). For the sake of qualitative analyses, the data set also records place names to describe the footprints. All place names are pre-defined and correspond to different levels of administrative units. From the data set, we identified 370 places in total, including the 4 Direct-Controlled Municipalities (Beijing, Shanghai, Tianjin, and Chongqing), Macau, Hong Kong, 332 prefecture-level units[3], 13 county-level units, and 19 cities in Taiwan. The administrative division system of China is rather complicated and the readers may refer to a Wikipedia entry[4] and a webpage in Chinese[5] for a better understanding. In this research, a place is abstracted to a point that is the capital city's (or town's, in very rare cases) location of the units. All check-in locations inside the place are captured to the point so that we can investigate the aggregate level of spatial interaction. For simplicity, we use the term "city" for a cluster of check-in points. It is natural the total check-ins of a city is positively correlated with its size. This is confirmed by the distribution of check-ins (Figure 1B), which is consistent with the rank size distribution of Chinese cities [56].

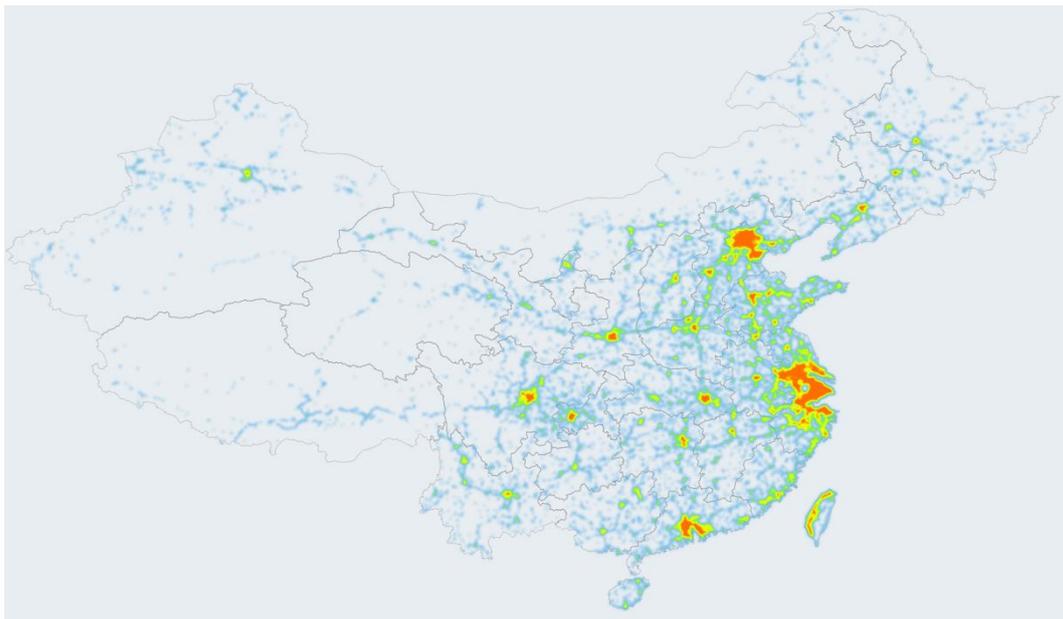

(A)

---

[2] We obtained the data set due to the collaboration between our laboratory (Geosoft@PKU) and the LBSNS provider.
[3] Up to 2013, there are 333 prefecture level units in China. In the data set, we do not find any check-in record inside Shigatse, Tibet.
[4] http://en.wikipedia.org/wiki/Administrative_divisions_of_the_People's_Republic_of_China
[5] http://www.gov.cn/test/2005-06/15/content_18253.htm

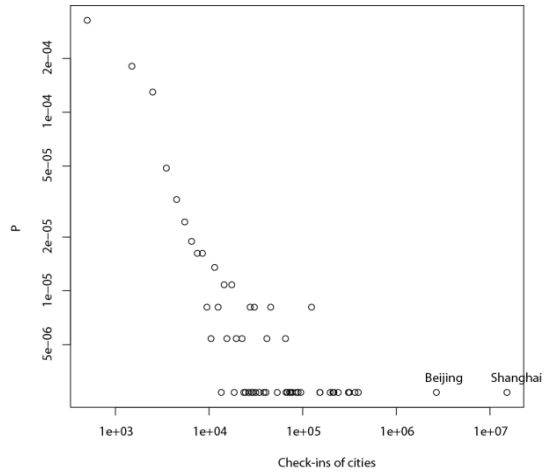

(B)

**Figure 1. Heat map of all check-in points and frequency distribution of check-ins in the 370 cities.** (A) The map created using density estimation clearly depicts the distributions of cities and transportation networks in China. Note that The South China Sea Islands are not shown for simplicity. (B) The frequency distribution exhibits a heavy tail characteristic. Shanghai and Beijing, the two biggest cities in China, have most check-in records.

Given a user, his (or her) trajectory can be formalized as $\{<City_1, T_1>, <City_2, T_2>, …, <City_n, T_n>\}$, where $n$ is the check-in number of the user, and the $City_i$ was visited at time $T_i$ ($1 \leq i \leq n$). Figure 2A plots the distribution of all users' check-in numbers, which follow a power law distribution well. From the footprints of each user, we can extract the cities that he (or she) visited. The distribution of visited cities is shown in Figure 2B also illustrates a heavy tailed distribution. Among all users, 237,000 (45.6%) individuals have visited at least two cities, and we can thus construct inter-urban scale trajectories for these users (Figure 2C).

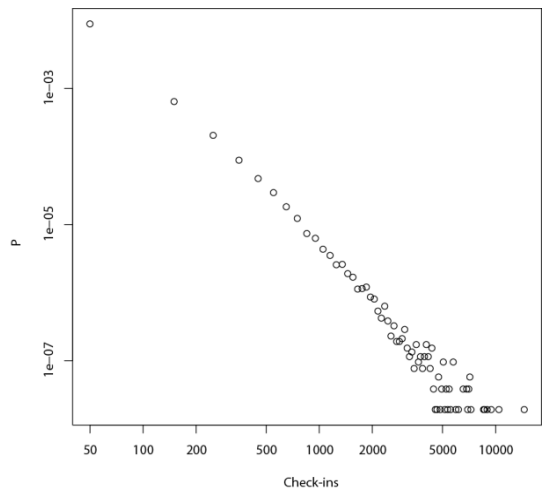

(A)

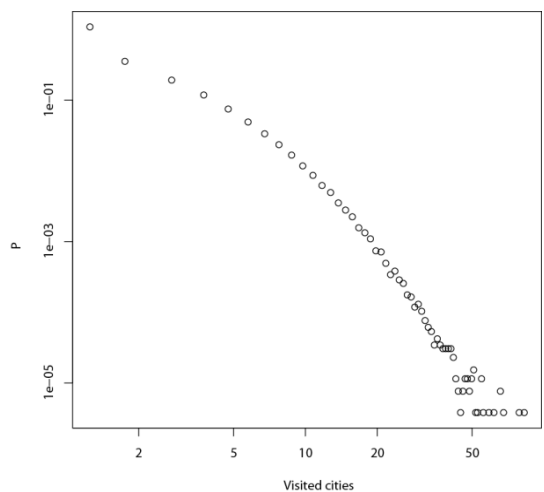

(B)

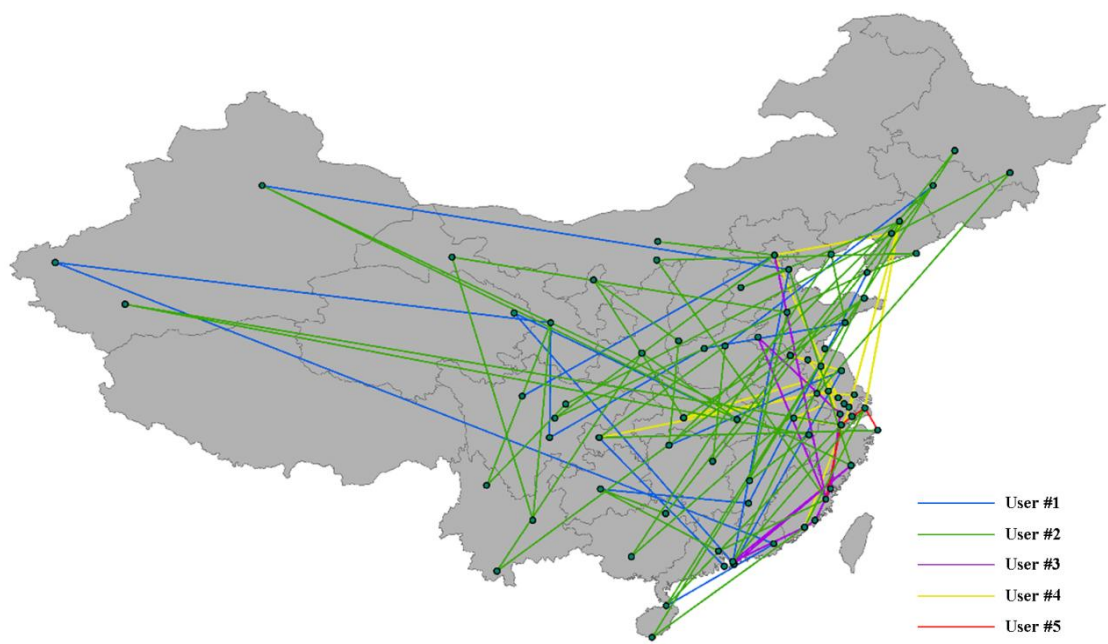

(C)

**Figure 2. Characteristics of check-ins from the perspective of users.** For each user, we compute the number check-ins, $N_h$, and the number of visited cities, $N_c$, so that the inter-urban movements can be extracted. Note that $N_h$ and $N_c$ are not well correlated, since a user may check in many times in the same city. (A) Frequency distribution of $N_h$. (B) Frequency distribution of $N_c$. One user visited 83 cities, which is the maximum of all users. (C) Five anonymous example individuals' trajectories.

## 3.2 Data evaluation: a comparative approach

The inter-urban movements extracted from check-in records are associated with representativeness issues. In other words, not all individuals are registered users of a LBSNS. According to the statistics of Foursquare[6], a large proportion of its registered users are young and the users are likely to check in at particular places such as airports. The same is true for the Jiepang data set. To evaluate data, we introduce the flight passenger data of year 2011 as a comparison. The flight data set includes 79 cities and 541 pairs of flows, denoted by $T_{fij}$ for cities $i$ and $j$. $T_{fij}$ is the number of passengers between cities $i$ and $j$ in 2011. From the check-in data, we also compute the trip numbers ($T_{cij}$) for the 541 city pairs. $T_{cij}$ and $T_{fij}$ are roughly positively correlated with a low $R^2$=0.533 (Figure 3A). The low $R^2$ indicates that the check-in records capture movements beyond the flight data but underestimate a number of flight trips. To investigate the first case, we calculate $T_{cij}/T_{fij}$ and select 50 pairs with the top highest $T_{cij}/T_{fij}$ (Figure 3B). Three facts are found in the 50 city pairs. First, the distances between 32 pairs of cities (64%), depicted in blue color in Figure 3B, are less than 1,000km. Within this distance interval, flight trips are not dominant and railway travels is a major competitor in China. Hence, we can obtain relatively more movements from check-in records than from flight data. Additionally, a direct flight line does not exist for very short distance city pairs, such as Beijing-Tianjin and Guangzhou-Shenzhen. The trips between these city pairs can be estimated by the check-in data. Second, as a new mobile application, a LBSNS has different acceptance rates across the country, depending on the regional ICT (information and communications technology) development level. Among the 50 city pairs, 44 pairs (88%) include Shanghai or Beijing. This can be attributed to the high ICT development level of the two cities and more registered users relative to the other cities. Last, it is interesting that the city list covers some China's top tourism destination places (e.g. Jiuzhaigou in Sichuan, Lijiang in Yunnan, Zhangjiajie in Hunan, Sanya in Hainan, and Guilin in Guangxi), indicating a person is more likely to check-in when he (or she) is on a tour. It is natural since one may be excited during a tour and want to share some new contents with his (or her) friends via a social media application. This leads to a high check-in probability and we can thus extract more trips. On the contrary, the low $T_{cij}/T_{fij}$ values can always be interpreted by either long trip distances or low ICT development levels. Note that we only introduce the flight data as a comparison due to the data limitation. If we have the flows of other transportation modes such as railway, similar results can still be obtained, that is, flows derived from different datasets are not well correlated and their ratios are influenced by the features of cities.

---

[6] http://www.factbrowser.com/tags/foursquare/

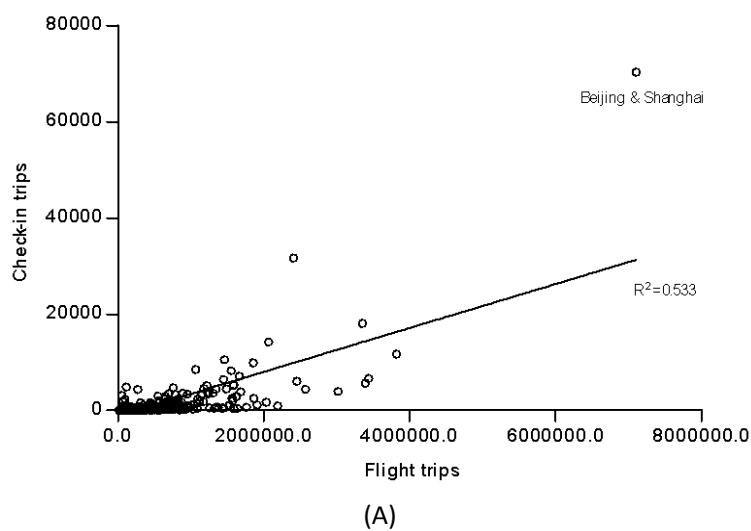

(A)

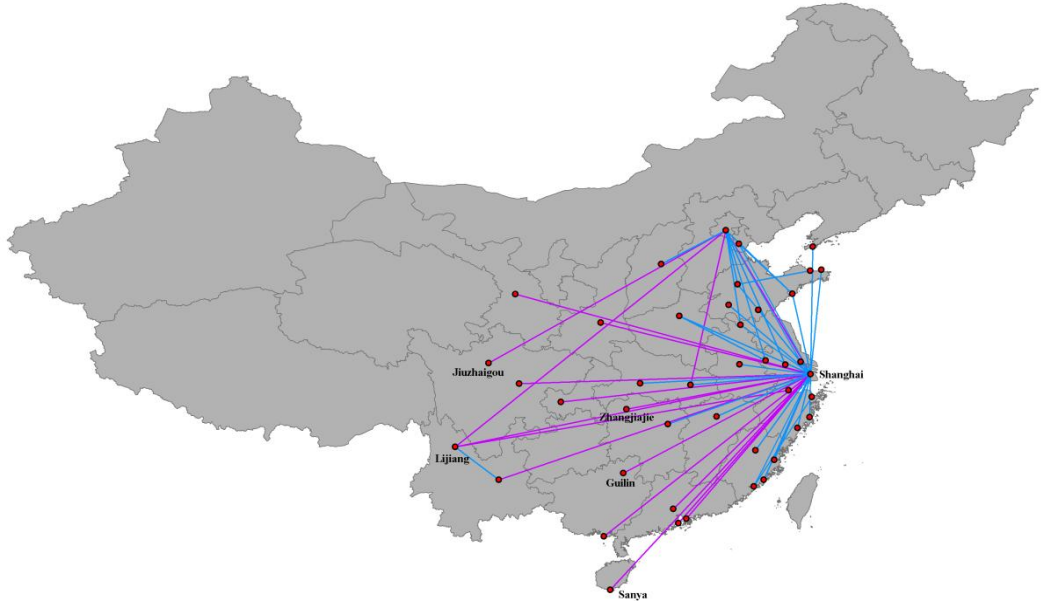

(B)

**Figure 3. Comparison between trips extracted from check-in records, denoted by $T_{cij}$, and flight trips $T_{fij}$.** (A) Scatter plot of $T_{cij}$ versus $T_{fij}$, indicating a weak positive correlation. (B) 50 city pairs with the top highest $T_{cij}/T_{fij}$.

## 4. Analyses

At present, most human mobility research is conducted based on a large population instead of a single person [7-9,30]. Hence, the movement frequency between two places can be used to measure the interaction strength between them. Distance has been widely accepted as an important factor in both individual movements [3,9,40] and collective spatial interactions [19-21]. There is little research on linking these two aspects. Hence, we construct a spatially-embedded interaction network and introduce the gravity model to quantify the distance impact behind the network and to examine whether the distance decay can reproduce the observed displacement distribution, which is critical in human mobility studies. Network science provides a new perspective to understand spatial interactions. Recently, much literature has introduced community detection methods to regionalize a study area [52,57]. The distance decay effect, however, were not considered in such studies, despite its importance in modeling spatial interactions. In the following sections, we focus on displacement distribution and community detection, two important topics in human mobility patterns and spatially-embedded networks, using two fundamental concepts in geographical analyses: spatial interaction and distance decay effect.

## 4.1 Fitting the gravity model

From the extracted trajectories, we can compute both the check-in number for each city and the movement between each two cities. An undirected weighted network, denoted by **G**, is constructed from the interaction strengths (Figure 4A). Note that the movements between cities are actually directed, and we sum the flows in two directions to represent the interaction strengths. **G** has 370 nodes and 15101 edges (graph density = 0.351). In terms of other statistics of **G**, the graph diameter is 3, the average degree $\langle k \rangle$=81.6, the average shortest path $\langle l \rangle$=1.781, and the average clustering coefficient $\langle C \rangle$=0.657. Compared with a random network, the relatively low $\langle l \rangle$ and high $\langle C \rangle$ suggest that **G** has properties of a small world network.

The edge weights follow a power law distribution (Figure 4B). It is similar to the spatial interaction distributions identified from different data sets [15,16,32]. Kang et al. argued that such a power law distribution mainly derives from the city size distribution, given that its distance decay effect is weak [15].

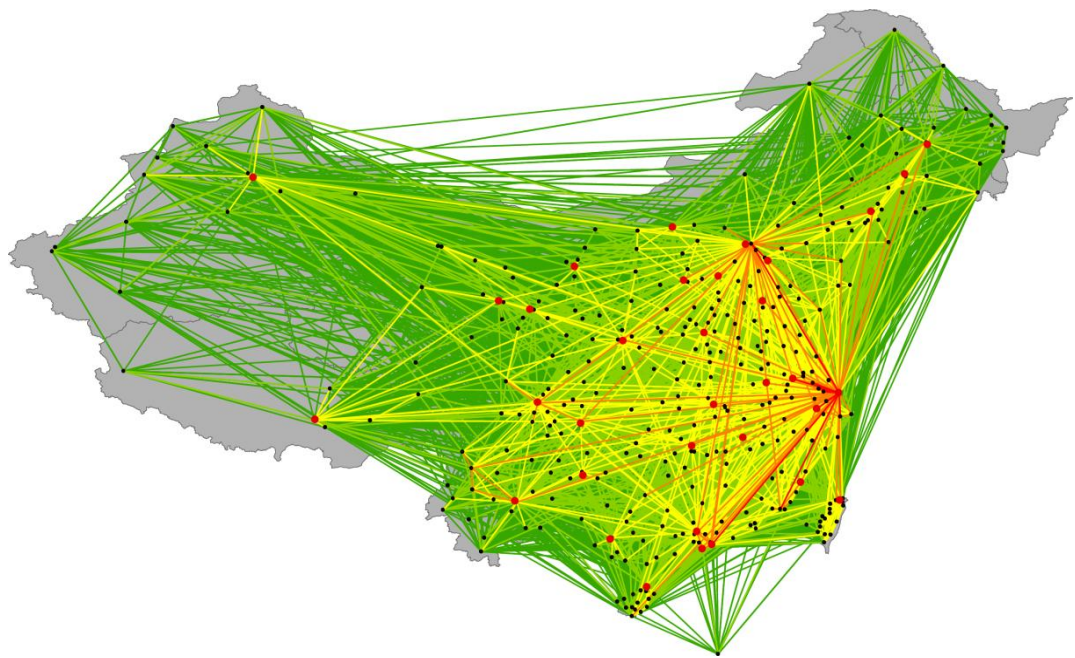

(A)

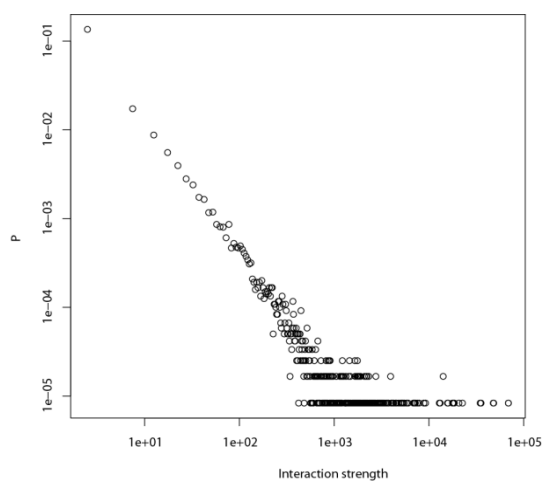

(B)

**Figure 4. Characteristics of interaction strengths between the 370 cities.** (A) Interaction map of the 370 cities. The red lines indicate stronger interactions. The maximum value is 137,847, which is the number of trips between Shanghai and Suzhou, extracted from the check-in data set. The red dots represent capital cities of provinces in China. (B) Frequency distribution of edge weights (or interaction strengths) between cities.

In this research, we quantitatively estimate the distance decay effect by fitting the gravity model. Because of the low graph density, we adopt the PSO method to find the best fit. According to the PSO method, we try different β values, from 0.1 to 2.0 with a step of 0.1, in the gravity model. The goodness of fit (GOF) is measure using the correlation coefficient between the observed and estimated interactions. For each fixed β value, say 1.0, the PSO method is used to search the best GOF, where each particle is a 370-dimensional vector denoting the theoretical sizes of all cities.

The maximum GOF=0.985 is achieved when β=0.8. The exponent is close to the value observed from air passenger flows in China [11] but lower than the distance parameters, which vary between 1.0 and 2.0, estimated using intra-urban movement data [9,30]. Figure 5 plots the relationship between the estimated interactions and real interactions between cities. The high GOF indicates that the inter-urban interactions are governed by the gravity model with a power law distance decay effect.

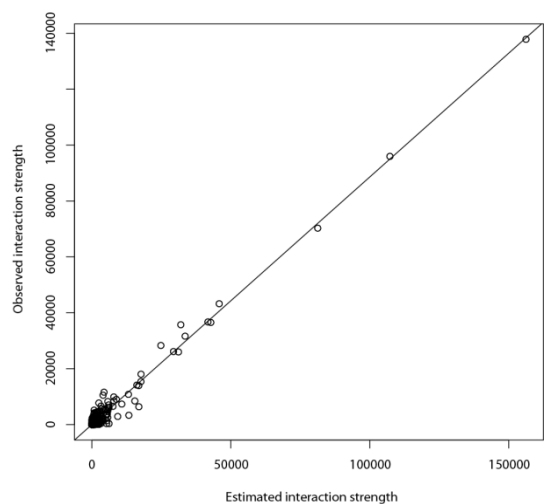

**Figure 5. Plot of estimated versus observed interaction strengths when β=0.8, indicating the observed inter-urban interactions can be well fitted using the gravity model.**

Some research has pointed out shortcomings of the gravity model [40,57]. They argued that the gravity model cannot well predict human movements at both the individual level and the collective level. Recently, Masucci et al. compared the gravity model versus the radiation model developed in Ref. [40] using several empirical data sets but did not reach a clear conclusion [58]. In these studies, populations are directly used in the gravity model as the nodes' sizes. For a place, however, its "mass" that leads to the observed interaction strengths is not necessary to be positively linearly correlated with its population. As shown in Figure 1b, the check-in number in Shanghai is much greater than that in Beijing, although the populations of the two cities are close. Hence, we suggest do not negate the gravity model easily. The appropriate way to adopt the gravity model is fitting the model according to the real interaction strengths and distances between places instead of predicting interactions directly based on populations or other similar attributes.

As pointed in Section 3.2, check-in data only partially capture inter-urban movements and there exist sampling biases. Sampling biases also exist in the air passenger data or even a data set collected based upon other transport modes (e.g. railway) as users of different modes are often correlated with their socio-economic attributes. A single data set represents one aspect of human trips and thus they might not be consistent with each other. It is interesting that the check-in data and the air travel data can be well fitted by different gravity models with different distance parameters and theoretical size sets. Figure 6 demonstrates a framework for integrating different interaction systems. Suppose we have $N$ places and $K$ interaction systems. Let $I_{kij}$ denote the interaction between $i$ and $j$ in the $k$th system and $d_{ij}$ is the distance ($i,j$=1,…,$N$ and $k$=1,…,$K$). Since $\{I_{kij}\}$ and $\{d_{ij}\}$ are known, ideally, we can get $K$ gravity models. For the $k$th model, the distance parameter $\beta_k$ and the theoretical sizes $\{\hat{S}_{k1}, ..., \hat{S}_{kN}\}$ can be estimated. Since a city plays different roles in various interaction systems, a comparative investigation on the derived theoretical size set $\{\hat{S}_{1i}, ..., \hat{S}_{Ki}\}, 1 \leq i \leq N$ helps us better understand the $i$th city. For example, in the flight network, the attraction of Beijing is a bit greater than that of Shanghai, according to Ref. [11]. With regard to the check-in data, on the contrary, Shanghai is much more important than Beijing. Such a difference is caused by the fact that Beijing is China's political center with more flight lines but Shanghai exceeds Beijing in both economy and ICT development. In conclusion, although the check-in data are biased, they still obey geographical laws and we can obtain particular findings from the data that might be hidden in other data.

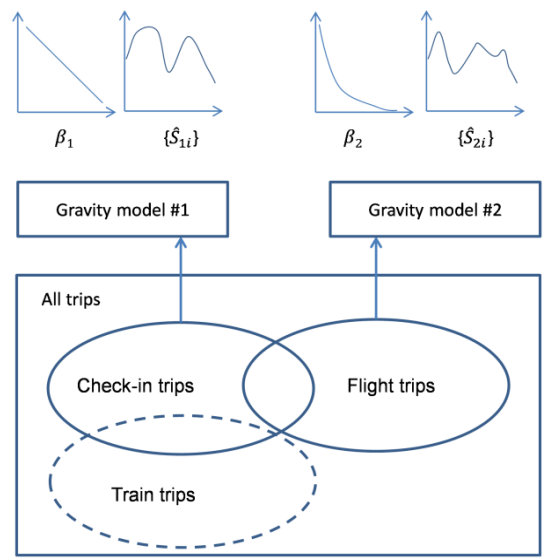

**Figure 6. Different data sets represent different aspects of "the ground truth" of human movements and thus can be used for revealing different roles of the same city.**

In terms of the human mobility pattern, the displacement distribution can be well fitted by an exponential distribution $P(\Delta d) \sim \exp(-\alpha \Delta d)$, where $\alpha$=0.002 and $\Delta d$ is measured in kilometers (Figure 7). It is interesting that the inter-urban displacement distribution do not have a heavy tail. Similar distributions have been observed using other data sets such as taxi trajectories [9,31] and mobile phone records [30]. Liu et al. attributed the observed distribution to the impact of geographical environments such as population distribution [9]. With regard to the inter-urban trips, the cities' locations and their populations will influence the displacement distribution.
The exponential displacement distribution is seemingly inconsistent with the power law distance decay, which implies a slower distance decay effect. Liu et al. [9] and Liang et al. [39] suggested that the observed displacement distribution can be well interpreted by integrating the inherent distance decay effect with geographical heterogeneity, and thus proposed a probability form of the gravity model:

$$Pr(T_{ij}) = \frac{kP_iP_j}{d_{ij}^{-\beta}} \qquad (2)$$

where $T_{ij}$ denotes the event that there is a movement between *i* and *j*. We adopt the estimated city sizes and distance decay parameter β=0.8 to randomly generate the same number of synthetic trips with the number of observed trips using the Monte Carlo simulation approach. The distributions of both observed and synthetic displacements are shown in Figure 7. We can see that the two distributions match well, further confirming the underlying gravity model.

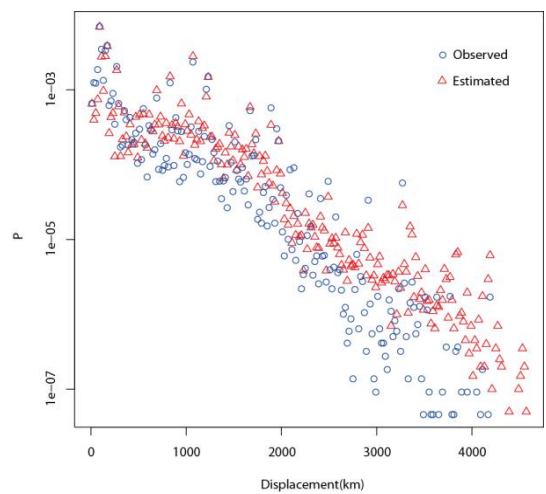

Figure 7. Displacement distributions of observed and estimated trips.

As mentioned earlier, inter-urban (or region) interaction is a traditional topic in geographical analyses, while recently human mobility patterns drew much attention with the availability of big trajectory data. This research indicates that the aggregate level of spatial interactions and individual level of movements can be viewed as two sides of the same coin. If the collective spatial interactions can be interpreted by the gravity model (Figure 5), then it is possible that the individual level movements are governed by the gravity model with an identical distance decay parameter (Figure 7). Note that there are some efforts to introduce the gravity model or similar models for mobility patterns. For example, Bazzani et al. proposed a chronotopic model that takes into account attractivity to simulate intra-urban movements [59]. Recently, Liang et al. proved that the exponential displacement distribution can be obtained from the gravity model [39]. Besides the gravity model, some models are built based on benefits (or opportunities) and thus distance plays an indirect role [3,38,40]. These models take into account the decision when individuals plan a trip to a random destination, such as finding a restaurant. However, many inter-urban movements, such as returning to hometown during holidays, are nonrandom so that the benefit based models do not apply.

It should be pointed that an ecological fallacy exists in extending collective level statistics to the individual level. Although various existing models, including the gravity model in this research, well reproduce the observed displacement distribution, it is still questionable that each individual's movements follow the same gravity model. Figure 8 demonstrates two extremely contrary cases with the same collective statistics. The data set contains four individuals' (denoted by A, B, C, and D) trajectories, the displacement distributions of which are represented using different colors. The first plot (Figure 8A) actually represents the model described by Equation 2, that is, each individual's movements exhibit a clear distance decay effect. However, we cannot deny the situation depicted in Figure 8B, where each individual moves with a roughly fixed distance, just like the commute trips inside a city. Most real individual level movements are the mixture of the two cases. We need further studies to decouple them using more detailed trajectory data sets.

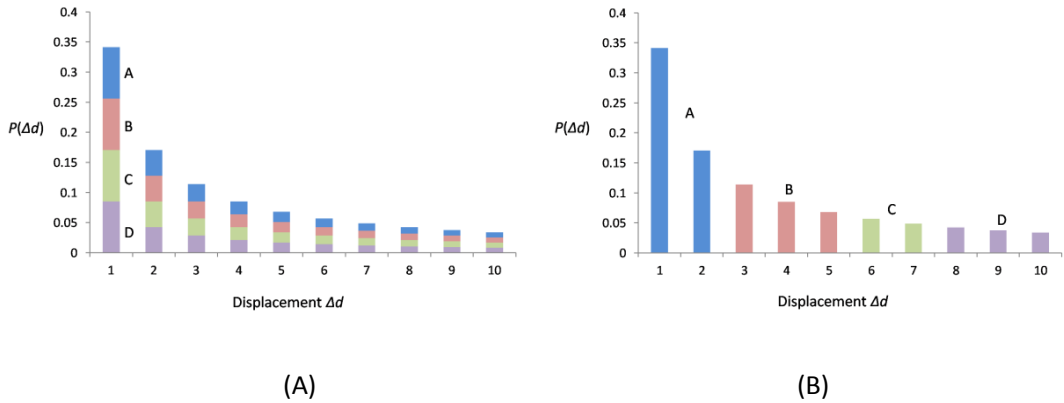

(A)                                  (B)

**Figure 8. Different individual level movement patterns may lead to the same collective statistics.**

## 4.2 Identifying network communities

For a spatially-embedded network, the community detection method can help us to reveal its structure. In this research, we create a Voronoi diagram based on the 370 cities and merge Voronoi polygons containing cities in the same community so that all communities can be spatialized and visualized. The multilevel algorithm developed by Blondel et al. [46] is adopted to optimize the modularity measure. Additionally, considering that most community detection algorithms are associated with randomness and thus we will get slightly different results if running the same algorithm for several times, the method proposed in Ref. [57] is adopted. We perform the algorithm for 20 times and the result is depicted in Figure 9, where the thicker borders indicate that they are boundaries in more resulting maps of community detection.

From Figure 9, we can find two facts about the partition result. First, all communities are spatially connected, although we do not impose the adjacency constraint during the procedure. Second, a number of communities roughly coincide with the administrative units, that is, the provincial units in China. Provinces such as Jilin, Henan, Guizhou, and Guangdong are clearly delineated in the resulting map. Note that the slight inconsistency between community boundaries and province boundaries is partially due to that the Voronoi polygon instead of the actual administrative area of a city is used to visualize the communities.

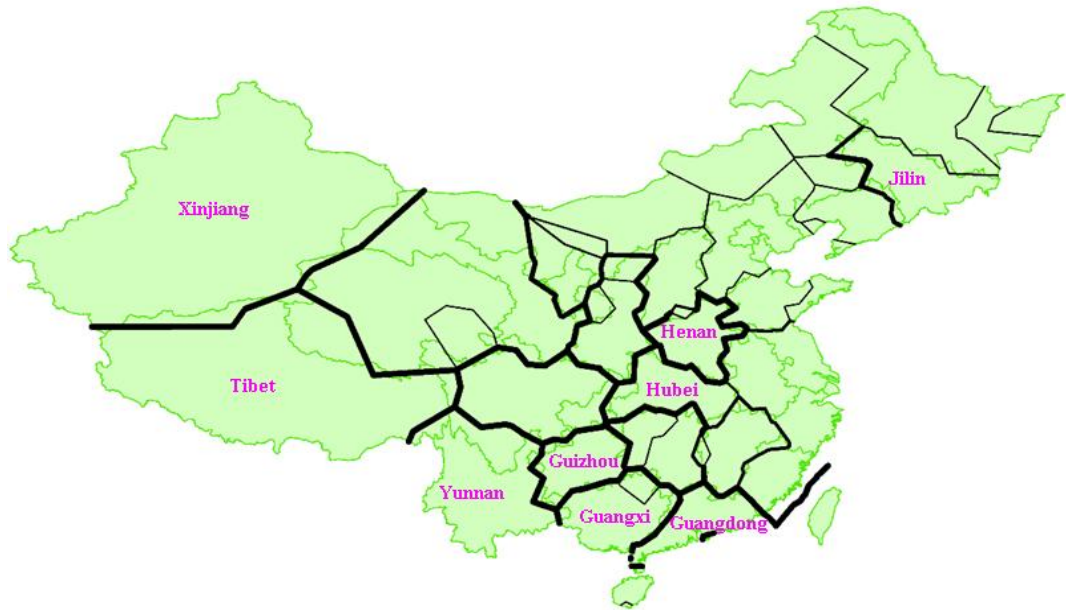

**Figure 9. Communities detected from the interaction network *G*.**

The partition pattern has been observed from various spatial networks [51-54]. The first feature, spatial connectedness, can be attributed to the distance decay effect in spatial interactions. Because of the distance decay, closer places generally have stronger interactions and thus are likely to be classified in to the same community. However, the second observation, i.e., coincidence of the identified communities with administrative units, has not been well interpreted yet. We suggest that the distance decay effect is different for intra-province trips and inter-province trips. The political characteristics of China make two cities inside the same administrative unit usually have a relatively stronger socioeconomic integration, indicating a high frequency of intra-province movements. In other words, the distance decay effect in intra-province trips is weaker than that in inter-province trips. Unfortunately, the number of intra-province city pairs (2053, about 1/7 of the total city pairs) extracted from the check-in data is small and cannot fit the gravity model very well. We simply redraw Figure 5 in a log-log plot and use different symbols to distinguish intra-province and inter-province interactions (Figure 10). It is clear that intra-province interactions are in general greater than inter-province interactions when compared with the estimated interactions computed from the gravity model, suggesting that the gravity model with β=0.8 underestimates the intra-province interactions and we should use a smaller exponent instead. In other words, administrative boundaries play a role of obstacle for human inter-urban movements and communications. This provides a reasonable explanation to the community detection result, as well as the findings reported based on different data sets [52,54].

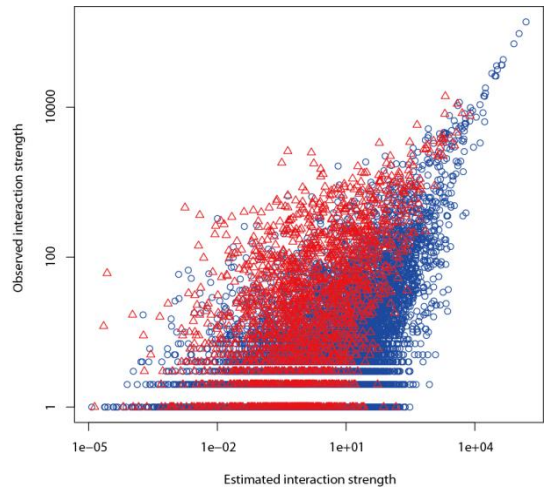

**Figure 10. Log-log plot of estimated versus observed interaction strengths when β=0.8.** The red triangles and blue circles represent interaction between cities in one province and two different provinces, respectively.

## 5. Discussion and conclusions

Human mobility patterns have been a hot research topic in many areas. However, existing studies do not differentiate movements at different spatial scales. Particularly, due to the data limitation, little literature investigated nationwide inter-urban trips. For the first time, this research adopts the check-in data to analyze the inter-urban movements. Our findings include the following four aspects. First, the inter-urban displacements follow an exponential distribution and do not have a heavy-tail property. This distribution is similar to that observed in intra-urban movements. Liu et al. suggested that the geographical environment is a reason for the thin tail in intra-urban displacement distributions. For inter-urban trips [9], this impact still exists. If all cities in this research are identical and have the same mass value in Equation 2, the movements will obviously follow a power law distribution. It is the size and location characteristics of all cities that lead to the difference between power law distance decay and exponential displacement distribution.

Second, the spatial interactions reflected by the check-in data can be well fitted by the gravity model. This confirms again the power law distance decay effect in spatial interactions, which has been found from many different data sets. Some existing research argued that the gravity model cannot well predict the spatial interactions if the place populations are directly used as the masses in the model. This research, on the contrary, illustrates that fitting the gravity model to estimate both the places' theoretical sizes and the distance decay function is an appropriate approach.

Third, this research points out the connection between spatial interactions and human mobility patterns. The distance decay function $d^{-\beta}$ can also be used to interpret individuals' movements. The distance parameter β=0.8 is less than those estimated from intra-urban movements, indicating a weaker distance decay effect. We also clarify the ecological fallacy issue in modeling human mobility patterns. Hence, a safe statement is that we "cannot reject" an individual level model if the statistics of the synthetic trajectories generated based on the model match the observed statistics. To construct a precise individual level model needs long-term and detailed trajectory data.

Last, by constructing a spatially-embedded network from the check-in data, we regionalize China's territory using a community detection method. The result exhibits similar patterns to previous studies, that is, most communities are spatially connected and coincide with geographical units, which are provinces in this research. Such patterns can also be attributed to the distance decay effect that makes closer cities in general have stronger connections and thus be clustered together. We also find a difference between the distance decay effects in intra-province and inter-province trips. It is the difference that makes interactions between cities in the same province relatively stronger and classified into the same community.

Human mobility patterns and spatially-embedded networks have drawn much attention in recent complexity science studies, where much literature focuses on finding the underlying geographical impacts. Meanwhile, spatial interactions in different spatial scales are widely investigated in geographical analyses. Distance obviously plays an important role in human mobility patterns, spatial interactions, and spatially-embedded networks. The distance decay effect decreases the probabilities of long-distance movements as well as the interaction strengths between faraway places, and consequently shapes the topological structures of spatial networks. Based on an empirical data set, this research makes an initial effort to bridge the three concepts using the distance effect. Inversely, with the rapid development of complexity science, human mobility patterns and spatially-embedded networks provide a new perspective and new tools to revisit conventional geographical analyses. It is more valuable in the era of big data since it becomes easier for us to collect various data for representing movements, measuring interactions, and constructing spatial networks.

# Acknowledgments

We thank F. Wang, D. Tong, and L. Yin for useful comments, and J. Wang for providing the flight passenger data.